\documentclass[sigconf]{acmart}

\usepackage{enumitem}
\usepackage{booktabs}
\usepackage{xcolor}

\AtBeginDocument{%
  \providecommand\BibTeX{{%
    \normalfont B\kern-0.5em{\scshape i\kern-0.25em b}\kern-0.8em\TeX}}}

\setcopyright{none}

\acmConference[CARRV 2020]{Fourth Workshop on Computer Architecture Research with RISC-V}{May 29, 2020}{Virtual Workshop}
\acmYear{}

\newcommand{\code}[1]{ {\tt{#1}}}

\begin{document}

\title{Enabling Virtual Memory Research on RISC-V with a Configurable TLB Hierarchy for the Rocket Chip Generator}


\author{Nikolaos Charalampos Papadopoulos}
\authornote{Contact author}
\affiliation{\institution{National Technical University of Athens}}
\email{ncpapad@cslab.ece.ntua.gr}

\author{Vasileios Karakostas}
\affiliation{\institution{National Technical University of Athens}}
\email{vkarakos@cslab.ece.ntua.gr}

\author{Konstantinos Nikas}
\affiliation{\institution{National Technical University of Athens}}
\email{knikas@cslab.ece.ntua.gr}

\author{Nectarios Koziris}
\affiliation{\institution{National Technical University of Athens}}
\email{nkoziris@cslab.ece.ntua.gr}

\author{Dionisios N. Pnevmatikatos}
\affiliation{\institution{National Technical University of Athens}}
\email{pnevmati@cslab.ece.ntua.gr}



\keywords{RISC-V, Rocket Chip Generator, TLB, Memory Management Unit}

\renewcommand{\shortauthors}{N. C. Papadopoulos, V. Karakostas, K. Nikas, N. Koziris, D. N. Pnevmatikatos}

\begin{abstract}

The Rocket Chip Generator uses a collection of parameterized processor components to produce RISC-V-based SoCs. It is a powerful tool that can produce a wide variety of processor designs ranging from tiny embedded processors to complex multi-core systems. In this paper we extend the features of the Memory Management Unit of the Rocket Chip Generator and specifically the TLB hierarchy. TLBs are essential in terms of performance because they mitigate the overhead of frequent Page Table Walks, but may harm the critical path of the processor due to their size and/or associativity. In the original Rocket Chip implementation the L1 Instruction/Data TLB is fully-associative and the shared L2 TLB is direct-mapped.  We lift these restrictions and design and implement configurable, set-associative L1 and L2 TLB templates that can create any organization from direct-mapped to fully-associative to achieve the desired ratio of performance and resource utilization, especially for larger TLBs. We evaluate different TLB configurations and present performance, area, and frequency results of our design using benchmarks from the SPEC2006 suite on the Xilinx ZCU102 FPGA.


\end{abstract}

\maketitle

\section{Introduction}
\label{sec:introduction}

Rocket Chip Generator (RCG) is a tool that uses the open RISC-V ISA to produce configurable SoCs. RCG supports fully-fledged Unix-like operating systems, and features important RISC-V extensions and accelerators. RCG is designed to target a wide range of application domains, ranging from embedded up to complex and multicore systems. 
To support this wide range of application domains, most of the processor components have been implemented as configurable templates in the Chisel high-level hardware construction language (HCL).
However, some of the Rocket Chip Generator components are still missing support for configurability. In this paper we focus on the Memory Management Unit (MMU) and specifically on the Translation Lookaside Buffer (TLB) hierarchy that lacks such configurability support. 
TLBs are essential in terms of performance because they mitigate the overhead of frequent page table walks, but may harm the critical path of the processor due to their size and/or associativity. Furthermore,
a configurable TLB hierarchy might be useful for performance scaling for faster processors such as the out-of-order BOOM \cite{Celio:EECS-2015-167}.

In the original Rocket Chip implementation only the number of TLB entries is configurable; the L1 Instruction and Data TLBs can only be fully-associative and the shared L2 TLB direct-mapped. 
However, that approach is not optimal for applications with large memory footprints that require larger TLB reach with many entries because 
(i) increasing the number of the fully associative L1 TLB may  increase the processor critical path and can impact the operating frequency of the entire design, and
(ii) a direct-mapped L2 TLB can experience many conflict misses, leaving significant room for  application performance improvement with the use of increased associativity. 
Clearly, this lack of configurability in the TLB may limit the efficient applicability of Rocket Chip SoCs for applications with large memory footprints that stress the TLB hierarchy.

In this paper we lift these restrictions and design and implement
configurable, set-associative L1 and L2 TLB templates that can
create any organization from direct-mapped to fully-associative to
achieve the desired ratio of performance and resource utilization,
especially for larger TLBs.
Furthermore, we modify existing replacement
policies to be compatible with our design, offering flexibility for performance and resource usage trade-offs.

We modify the L1 and L2 TLB mechanisms and specifically how TLB lookups, refills, 
flushes, and replacements are handled. Chisel 
allows the programmer
to produce circuit generators that are easily configurable. With our approach, just by adjusting the number of the sets and the ways of the L1/L2 TLB, all the TLB circuitry is properly configured. Corner cases such as direct-mapped and fully-associative organizations
are included, and the design is tailored to remove unnecessary components for
these cases. For example, if a direct-mapped organization is selected there 
is no need for replacement policy, so our Chisel code removes it altogether.

We use different L1/L2 TLB configurations to evaluate our design with benchmarks from the SPEC2006int suite \cite{spec01}.
We show that the largest evaluated TLB configuration improves performance by up
to 15.4\%,
with minimal impact in area and frequency.

In summary the main contributions of this paper are:
\begin{itemize}
    \item We implement a fully configurable Instruction/Data L1 TLB and shared L2 TLB that can output any design from direct-mapped to fully-associative, lifting the initial restrictions of configurability only by the number of entries. This leads to better scaling of performance and resources, especially for large TLBs. 
    We make our design publicly available\footnote{Available at https://github.com/ncppd/rocket-chip} to enable further research on the active topic of virtual memory support for RISC-V.
    \item We present a case study in which we evaluate the performance and resource usage of the Rocket Chip \cite{asanovic01} processor with different TLB configurations, by running benchmarks from the SPEC2006int \cite{spec01} suite on the Xilinx ZCU102 FPGA.
\end{itemize}

\section{Background}
\label{sec:background}

Here we provide information on virtual memory, the Chisel hardware description language, and the Rocket Chip Generator.

\subsection{Virtual Memory}

Virtual memory is an essential concept for processor design because it provides the illusion of a very large and private address space to each process running in the system. Virtual memory offers security through process isolation and also benefits programmer productivity since the operating system manages the memory mappings and the hardware accelerates the translations. 

RISC-V supports different Virtual Memory systems depending on the size of the address space (e.g. RV32 Sv32, RV64 Sv39/Sv48 \cite{waterman02}), in this paper we focus on RV64 Sv39 (39-bit address space) which supports 4KB base pages but also 2MB, 1GB super pages; the page table, that stores the memory mappings of each process, is implemented as a multi-level radix tree (3-level page table in RV64 Sv39).
A processor register called \code{SATP} (Supervisor Address Translation and Protection register) holds the root of the page table. The physical address is obtained after performing a sequential lookup in each page table level. The page table walker (PTW) that performs the virtual-to-physical address translations is typically implemented in hardware for improved performance.

\begin{figure} 
\centering
\includegraphics[width=0.4\columnwidth]{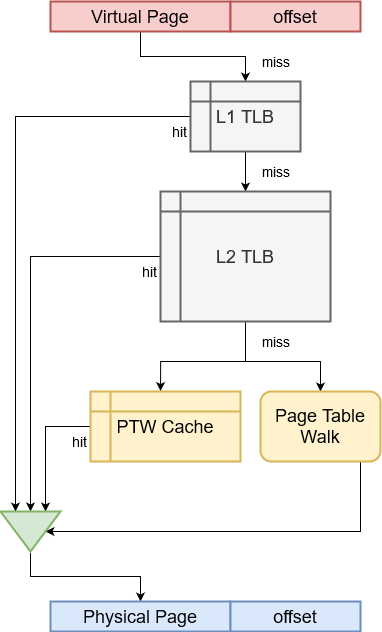}
\vspace{-5pt}
\caption{Overview of the MMU in Rocket Chip Generator.}
\label{figure_mmu}
\vspace{-5pt}
\end{figure}

To accelerate address translation without accessing the page table on every memory reference, a Translation Lookaside Buffer (TLB) is used which keeps the recently used translations. The TLB lies on the critical path of the processor and as a result its size and associativity are essential for the overall performance. To overcome this problem without sacrificing the hit rate, multi-level TLB organizations are used; the first level TLB (L1) is usually small (32-128 entries) but very fast, while the second level TLB (L2) is usually larger (128-1024 entries) but slower. Finally, a Page Table Walk cache is usually implemented to hold non-leaf intermediate translations of the page table to avoid searching levels of the page table (TLBs hold the leaf translations). Figure \ref{figure_mmu} shows these structures.

\subsection{Chisel}

Chisel \cite{bachrach} is a high-level Hardware Construction Language (HCL) embedded in the Scala language. Chisel enables the design of powerful circuit generators by utilizing Scala's high-level programming concepts like object-orientation, functional programming, parameterized typed and type inference. Chisel can generate synthesizable Verilog for both FPGA simulation and ASIC implementation. It can also output cycle-accurate C++ simulators which are very useful for hardware simulation and debugging.

\subsection{Rocket Chip Generator}

The Rocket Chip Generator (RCG) \cite{asanovic01} generates RISC-V ISA \cite{waterman01, waterman02} based systems using Chisel. It can also be considered as a library of processor parts that can easily be reused with any design written in Chisel. By default, the Rocket Chip Generator instantiates Rocket, an in-order core implementation, but also supports various core implementations including the BOOM out-of-order processor \cite{Celio:EECS-2015-167}. Rocket is a simple, 5-stage, in-order processor that implements the RISC-V ISA, including an MMU that supports page-based virtual memory, TLBs, instruction and data caches, and a frontend that features dynamic branch prediction with configurable sizes.

\section{TLB Hierarchy Design}
\label{sec:design}

In this section we provide an overview of the original implementation of the Instruction/Data L1 and shared L2 TLB in the Rocket Chip Generator. Then, we present the design and implementation of our proposed configurable L1 and L2 TLB. Our design can output any organization ranging from direct-mapped up to fully-associative TLBs. 

\subsection{Original TLB overview}
\label{sec:design:original}

\begin{figure} 
\centering
\includegraphics[width=0.8\columnwidth]{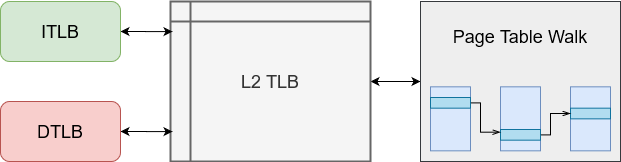} 
\vspace{-5pt}
\caption{Rocket Chip MMU organization.}
\label{figure_rocket_mmu}
\vspace{-8pt}
\end{figure}

Each processor has its own TLB hierarchy, as shown in Figure \ref{figure_rocket_mmu}.
The L1 Instruction and Data TLB hold address translations for the process code and the process data respectively.
The L1 Instruction/Data TLBs are built based on the same Chisel template in the RCG and only have minor differences regarding access privileges to pages. The L2 TLB is shared among the L1 Instruction/Data TLBs and can contain both Instruction and Data page translations. 

\subsubsection{L1 TLB}
The L1 Instruction/Data TLB stores the page translations in registers using a vector of \code{Reg} elements which create an array of positive-edge-triggered registers that output a copy of the input signal delayed by one clock cycle, depending on its activation signal. The original L1 TLB is fully-associative with configurable number of entries and uses a Pseudo-LRU Replacement Policy. The L1 TLB responds with a hit/miss indication on the next cycle and stores virtual-to-physical page translations of 4KB pages but also 2MB/1GB super pages.
 
\subsubsection{L2 TLB}
The Chisel template for the Page Table Walker (PTW) incorporates the shared L2 TLB. The PTW is connected with the L1 Instruction and Data TLBs though a Round-Robin Arbiter that selects the target virtual address to be translated. 
The shared L2 TLB is direct-mapped with configurable number of entries. Because of the direct-mapped organization  there is no need for a replacement policy. 
The L2 TLB stores the page translations using Chisel's \code{SyncReadMem/SeqMem} construct, which can be synthesized to FPGA Block RAM or ASIC SRAM. \code{SyncReadMem} basically creates a synchronous-read, synchronous-write memory, in this case with one read and one write port. Because of the \code{SyncReadMem} construct data are fetched on the next cycle; \code{SyncReadMem} outputs to a register with a purpose of performing a synchronous read operation. In order for the L2 TLB to sync with the rest of the PTW mechanism there are intermediate stages until the L2 TLB informs for a hit or miss. 


\subsubsection{Page Table Walk Cache}
The PTW Cache is a small fully-associative cache that stores the non-leaf virtual-to-physical page translations. In this paper we focus on the TLBs and leave the PTW Cache for future work.

\subsubsection{Limitations}

In the original Rocket Chip implementation, only the number of TLB entries is configurable.
However, that approach is not optimal for applications with large memory footprints that require larger TLB reach.
Increasing the number of the fully associative L1 TLB significantly increases the critical path of the processor and can impact the operating frequency of the entire design.
This happens because fully associative TLBs are typically implemented as CAMs. However, CAMs are a resource- and power-hungry structures, in both ASICs and  FPGAs \cite{wong01}.
Considering this, the original fully-associative L1 TLB is constrained and does not scale with application requirements. 
Increasing the size of L1 TLBs at lower associativity may increase the TLB reach and reduce the number of TLB misses without affecting the the overall resource usage/frequency.

Furthermore, because the L1 TLBs need to be fast, they are implemented using discrete registers that are generally precious resources both for ASIC and FPGA implementations. 
To mitigate the miss overhead of a relatively small L1 TLB, a larger but slower L2 TLB is introduced that stores translations in FPGA Block RAM or ASIC SRAM.
However, a direct-mapped L2 TLB can experience many conflict misses. In addition, L2 TLB misses are even more costly than L1 TLB misses, because they are resolved through page walks that incur increased latency. Associativity may reduce the number of conflict misses and improve the application performance.

To summarize, this lack of configurability in the TLB may limit the applicability of Rocket Chip Generator for workloads with large memory footprints that stress the TLB hierarchy.

\subsection{Configurable L1 TLB Architecture}
\label{sec:design:L1}

To develop a configurable L1 TLB we must consider a set of factors and  trade-offs. More specifically, the configurable Instruction/Data TLB should use registers using Chisel's \code{Reg} element for fast lookup time.
In addition, the configurable Data/Instruction TLB should be built by the same Chisel template with minor differences regarding the access privileges as mentioned earlier. Our implementation adheres to the aforementioned requirements and is compatible with the original implementation.
Next we describe how lookups, refills, replacements, and flushes are handled in our configurable L1 TLB.

\subsubsection{Lookup}
Whenever an address translation is requested, we obtain a \textit{tag} and an \textit{index} by splitting the \textit{VPN}. Using the \textit{index} we locate the target set and perform there a fully-associative search that matches the \textit{tag}. We modify the valid bit array and construct it as a \code{Vec} of registers, so every set has its respected valid bit array and can address it using the \code{index}.

\begin{table*}[htbp]
\centering
\begin{tabular}{ c c c c c c c } 
\midrule
\textbf{Conf. No} & \textbf{DTLB}  & \textbf{ITLB} & \textbf{L2 TLB} & \textbf{DTLB Reach} & \textbf{ITLB Reach} & \textbf{L2 TLB Reach}  \\ 
\midrule
\textbf{I} & fully-assoc., 32 entries & fully-assoc., 32 entries 
         & - & 128KB & 128KB & - \\ 
\textbf{II} & fully-assoc., 32 entries & fully-assoc., 32 entries 
        & 4-way, 128 entries & 128KB & 128KB & 512KB \\
\textbf{III} & fully-assoc., 32 entries & fully-assoc., 32 entries 
        & 4-way, 512 entries & 128KB & 128KB & 2MB \\
\textbf{IV} & 8-way, 64 entries & 8-way, 128 entries  
        & 8-way, 1024 entries & 256KB & 512KB & 4MB \\
\textbf{V} & 8-way, 128 entries & 8-way, 64 entries 
        & 8-way, 1024 entries & 512KB & 256KB & 4MB \\
\midrule
\end{tabular}
\caption{Rocket Chip L1 Instruction/Data TLB and shared L2 TLB configurations (Associativity /Size).}
\vspace{-20pt}
\label{tableConf}
\end{table*}

\subsubsection{Refill}
When a TLB refill is requested, we locate the target set that the virtual/physical address must be inserted using the \textit{index}. In case the set is not full, we select the first free slot. Otherwise, if the set is full we perform a Pseudo-LRU replacement. 

\subsubsection{Replacement Policies}
We modify the existing pseudo-LRU replacement policy and implement a set-associative alternative that uses the \code{Reg} construct. Support for a random replacement policy is already provided. A random replacement policy is an attractive alternative option thanks to its simplicity and can be also applied to TLBs; however, it may increase the TLB miss rate and hence degrade performance. 

\subsubsection{Flushing the L1 TLB}
When the OS modifies the page table, the stale TLB entries must be flushed. This happens when the OS executes the \code{sfence.vma} instruction to invalidate an entry. Using the \text{index} we retrieve the set that includes the entry to be flushed and perform a fully-associative lookup within that set using the \textit{tag}. The flushing of the TLB is done by zeroing the valid bit of the specified entry. 

\subsubsection{Limitations}
We initially developed the configurable L1 TLB in an older Rocket Chip edition that supported both base and super page sizes in the same TLB.
A constraint of a set-associative TLB structure that we must address concerns the page size: when the page size is unknown it is difficult to determine the least significant bits of the VPN in order to select a set \cite{talluri01, talluri02}. Therefore, we select to implement a configurable L1 TLB only for 4KB fixed page size. We also ported the configurable L1 TLB in a recent edition of the Rocket Chip in which this restriction is lifted, the TLB mechanism is separate for base/super pages, so our implementation of the configurable L1 TLB does not affect the superpage mechanism. More details about the Rocket Chip versions/commits that we modified are presented in Section~\ref{sec:methodology}.

\subsection{Configurable L2 TLB Architecture}
\label{sec:design:L2}

The L2 TLB was originally direct-mapped, an organization very simple in terms of replacement policies and TLB flushing. An address translation maps only to a unique TLB entry and as a result there is no need for a replacement policy. The \textit{valid bit array} is kept in register banks and not in the \code{SyncReadMem} that the TLB entries are stored. Obtaining a value from a register bank is completed in the same cycle in contrast with the \code{SyncReadMem} that has a cycle delay. As a result the valid bit array of the L2 TLB can be read and updated on the same cycle. This has the benefit of manipulating the valid bit without accessing the TLB array. The valid bit array is constructed as a \code{Vec} of registers the same way as in the L1 TLB.

\subsubsection{Lookup}
The L2 TLB lookup mechanism is similar to that of the L1 TLBs. The only difference is that the lookup in the L2 TLB introduces additional cycle delay due to the \code{SyncReadMem} construct. As a result we use registers to hold intermediate state.

\subsubsection{Refill}
In case of a refill, the L2 TLB handles it similarly with the L1 TLB. The only difference is the use of \textit{masks} to update a specific way in a set. Masks are a feature of the \code{SyncReadMem} construct to ease updating specific indexes inside a set.

\subsubsection{Replacement Policies}
To choose a replacement policy we must make a trade-off between area and performance. The pseudo-LRU replacement policy must keep track of the way access history and as a result impacts the total area when the TLB is large. On the other hand, using a random replacement policy has a nearly zero impact on the total area but may degrade performance. We implement both replacement policies for the L2 TLB. In Section~\ref{sec:evaluation} we choose to evaluate our set-associative design with the random replacement policy in favor of area constraints.

\subsubsection{Flushing the L2 TLB}
Flushing a TLB entry on a set-associative organization means that the entry must be located inside the selected set. In order to fetch the tags of the selected set there must be a cycle delay because of the \code{SyncReadMem} construct. To overcome this overhead and keep the flushing mechanism simple, we select to flush the whole set. Another approach would be to block the L2 TLB for one cycle to retrieve the set, and then flush the specific entry. We are considering implementing that in the future.

\section{Methodology}
\label{sec:methodology}

In this section we describe our evaluation methodology, including the hardware/software tools, metrics, and configurations. We initially developed the configurable TLB Hierarchy on an older Rocket Chip commit (7cd3352, April 3, 2018) that supported the Xilinx ZCU102 platform. Our contributions consists of about 80 and 70 lines of Chisel code added for the L1 and L2 TLB. Unfortunately that repository does not track the recent changes in the Rocket Chip Generator. As a result, we opted to use the old Rocket Chip version for our evaluation with the Xilinx ZCU102 platform. In addition, to ensure the relevance and compatibility of our approach with more recent versions of Rocket Chip, we ported our design to a more recent version (27120ee, Jan 22, 2020) that also features new mechanisms such as a sectored L1 TLB to further improve the TLB reach for 4KB pages, and a separate fully-associative L1 TLB for super pages. Our changes amount to about 50 and 70 lines of Chisel code for the configurable L1 and L2 TLB respectively in the recent version.  We validated our ported design using Verilator  simulations, and we plan to evaluate it on other supported FPGA platforms.


\subsection{Software and Hardware tools}

We follow a two-step process during the development of our TLB hierarchy. 
At first, we evaluate the L1/L2 TLB using Verilator \cite{verilator} 
to validate the 
correctness of our design and to remove any bugs; afterwards, we use the 
Vivado tools to compile our design for the Xilinx ZCU102 FPGA. 
In more detail, the development phase includes the following: 

\subsubsection{Verilator}
Verilator is an open-source tool that produces high-performance cycle-accurate C++/SystemC hardware models. Using \code{assert-printf} statements debugging becomes easier as Verilator produces logs of high verbosity. We use the official riscv-tests \cite{riscv-tests} as a sanity check,
and then orchestrate specific assembly tests that run upon the riscv-pk \cite{riscv-pk} (lightweight proxy kernel) which provides virtual memory support. Unfortunately, the downside of using Verilator is the slow emulation speeds in contrast with FPGAs.

\subsubsection{Software tools}
We use Sifive's Freedom-U-SDK \cite{sifive01} which sets up a minimal Linux environment. The Rocket Chip SoC boots the lightweight Buildroot \cite{buildroot} distribution on top of Linux kernel 4.15.0 with 4KB pages. We add new Buildroot packages that include simple TLB tests to verify that our design is working as expected, tools to retrieve performance counter results, and finally the SPEC2006 benchmarks \cite{spec01} (compiled using Speckle \cite{celio02}). We modify the Berkeley-Boot-Loader (BBL) \cite{riscv-pk}--which initializes machine registers and then boots the linux kernel--to set up several performance counters such as ITLB/DTLB and L2 TLB misses using the \code{mhpmeventXX} registers.

\subsubsection{FPGA Flow}
We use Vivado 2018.1 Design Suite for synthesis and placement. Vivado provides also results regarding resource usage. To evaluate the impact of the TLB hierarchy on application performance, we run a subset of the SPEC2006int \cite{spec01} benchmarks with different L1 Instruction/Data TLB and shared L2 TLB configurations. In all configurations we use a 4-way 32KB instruction cache and a 4-way 16KB data cache.

\subsection{Metrics and Benchmarks}
To evaluate our configurable TLB hierarchy we use the following metrics: (i) FPGA resource usage, i.e., flip-flops, look-up-tables (LUTs), and block RAM, (ii) TLB performance, i.e., TLB Misses-Per-Kilo-Instructions (MPKI), and
(iii) System performance, i.e., Instructions-Per-Cycle (IPC), a performance metric that isolates the impact of TLB implementation on the critical path, ignoring the processor frequency.
To evaluate the TLB and system performance we use benchmarks from the SPEC2006int suite \cite{spec01}. We use them with the test input set due to the limited physical memory (512MB) that our Xilinx ZCU102 platform exposes to the programming logic.

\subsubsection{Configuration scenarios}
We evaluate our configurable TLB hierarchy using different configurations for the L1 Instruction/Data TLB and shared L2 TLB. Table~\ref{tableConf} summarizes the evaluated configurations. We choose these configurations to cover a range of systems from small and embedded up to modern high-performance general-purpose systems. The TLB reach (i.e., number of entries $\times$ page size) covered by the L1 ranges from 128KB to 512KB, and for the L2 is up to 4MB. In the most lightweight configuration we choose not to include an L2 TLB to quantify the performance and area differences of the different configurations. Finally, in the most performant TLB configurations (Configurations IV, V) we swap the size of the Data and Instruction TLB to identify possible changes in performance without changing the L2 TLB. Note that in our evaluation we do not include a PTW Cache. Finally, the configuration scenarios are chosen to resemble well-known architectures:
\begin{enumerate}[label=\Roman*.]
\item Vanilla Rocket Chip without L2 TLB
\item Vanilla Rocket Chip including small L2 TLB
\item ARM Cortex A57 \cite{arm-cortex-a57}
\item Intel Skylake \cite{haswell}
\item Intel Skylake with swapped Instruction/Data TLB sizes. 
\end{enumerate}

\section{Results}
\label{sec:evaluation}

In this section we evaluate our configurable TLB hierarchy.
The purpose of our evaluation is twofold: (i) to show that the generated designs have minimal impact on area and frequency, and (ii) to show how TLB configurability 
affects performance.

\subsection{Area and Frequency Results}
\begin{figure}[ht] 
\vspace{-7pt}
\centering
\includegraphics[scale=0.6]{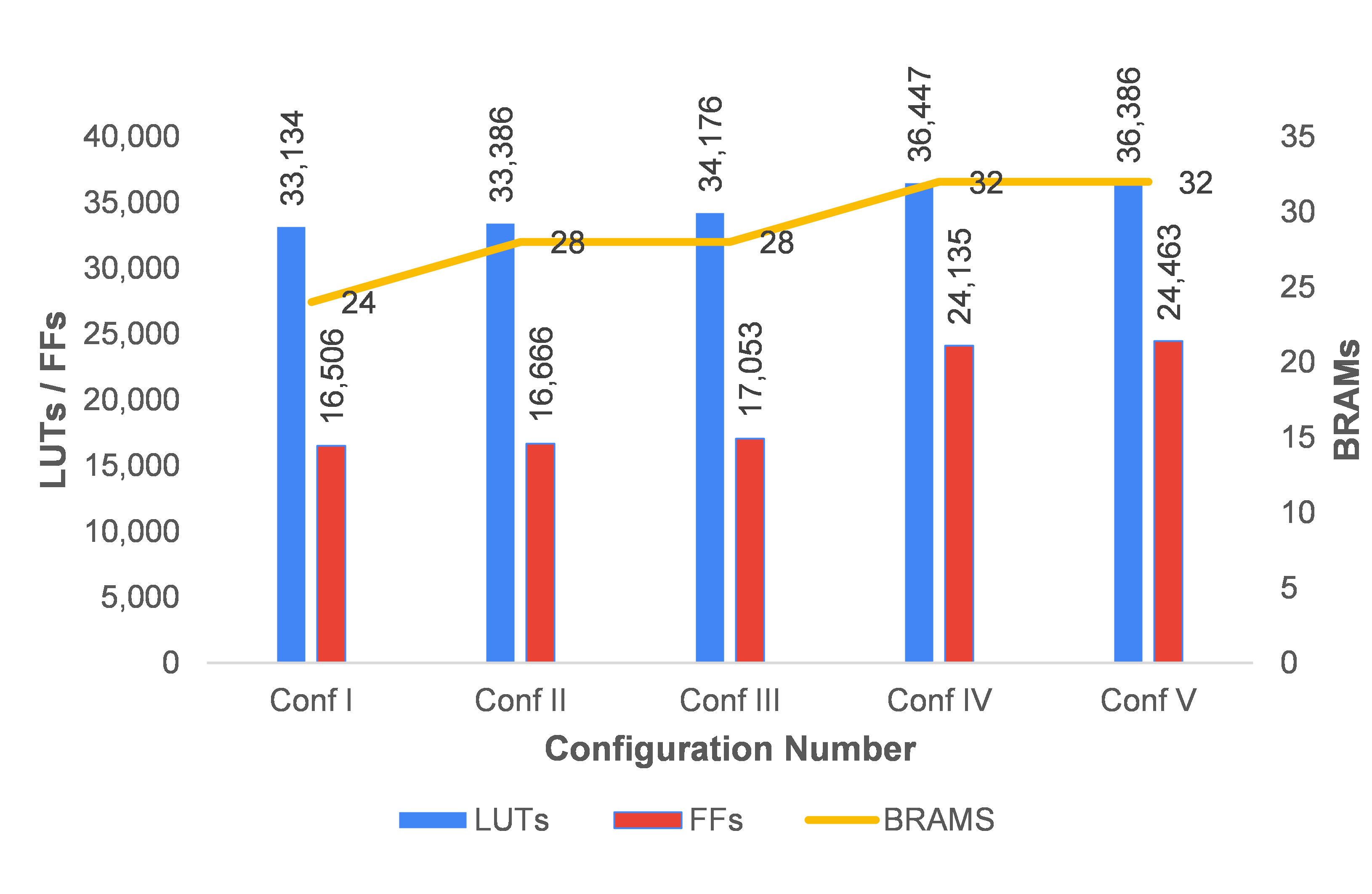} 
\vspace{-14pt}
\caption{Area results for different TLB configurations.}
\label{figure_area}
\end{figure}

Figure \ref{figure_area} shows the area results for the various configurations. We present the total area of the Rocket Chip SoC as reported by the Vivado 2018.1 Implementation stage. Note that the Instruction/Data L1 TLB structures use FFs and the shared L2 TLB uses BRAMs. 

\begin{table}[ht]
\centering
\vspace{-10pt}
\begin{tabular}{ l c c c c c } 
\midrule
\textbf{Configuration} & I & II & III & IV & V \\
\textbf{Frequency (Mhz)} & 189 & 187 & 186 & 188 & 186 \\
\midrule
\end{tabular}
\caption{Maximum operating frequency per configuration.}
\label{tableFreq}
\vspace{-20pt}
\end{table}

In the most lightweight scenarios (Conf I, II, II) Vivado 2018.1 reports that the full Rocket Chip SoC occupies 12\% of the total LUTs, 3\% of the total FFs, and 3\% of the total BRAMs of the Xilinx ZCU102.
Tuning up to the most performant configurations (Configuration IV, V) in terms of TLB hit rate, the Rocket Chip SoC occupancy increases to 13\% for total LUTs, and 4\% for total FFs/BRAMs. The FF usage is increased in Conf IV, V in order to accommodate the new TLB entries. 


Table~\ref{tableFreq} shows the maximum frequency achieved with all configurations.
The results show that the impact on the maximum operating frequency ranges from 0.53\%-1.59\%. In particular, Configuration IV has a 2$\times$ larger DTLB, 4$\times$ larger ITLB, and a 1024 entry L2 TLB, but exhibits only a 0.53\% drop in frequency compared to Configuration I. 

\subsection{Performance Results}

We now present the results of the SPEC2006int benchmarks that we obtained on the Xilinx ZCU102 FPGA board.

\begin{figure*}[] 
\centering
\includegraphics[scale=0.63]{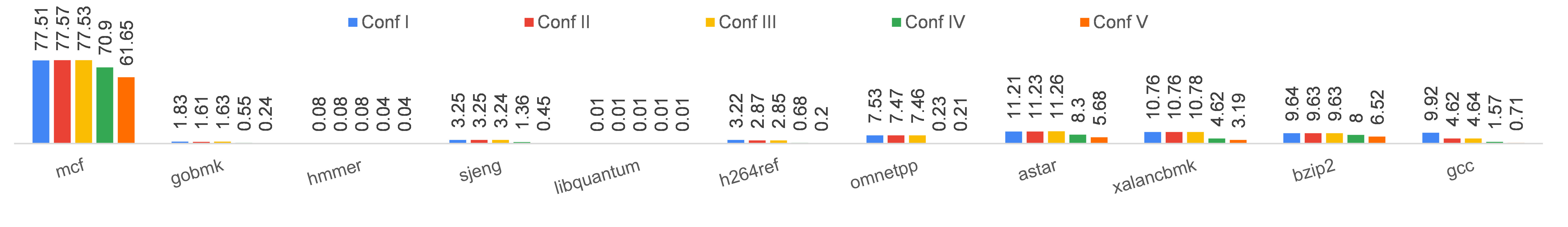} 
\vspace{-9pt}
\caption{Aggregated MPKI of the L1 Data/Instruction TLBs for the various TLB configurations.}
\label{figure_ditlb_mpki}
\end{figure*}

Figure \ref{figure_ditlb_mpki} shows the results of MPKI in the L1 Instruction/Data TLBs for the various configurations. We observe that gobmk, hmmer, sjeng and libquantum exhibit similar behavior in L1 TLB MPKI even with larger TLB configurations. The most demanding in terms of TLB miss rate is mcf, and even with the largest Configuration V the miss rate is still high. For Configuration IV - V the miss rate is nearly the same, with Configuration V performing better in all tests. Most misses come generally from the Data TLB.

\begin{figure*}[] 
\vspace{-7pt}
\centering
\includegraphics[scale=0.63]{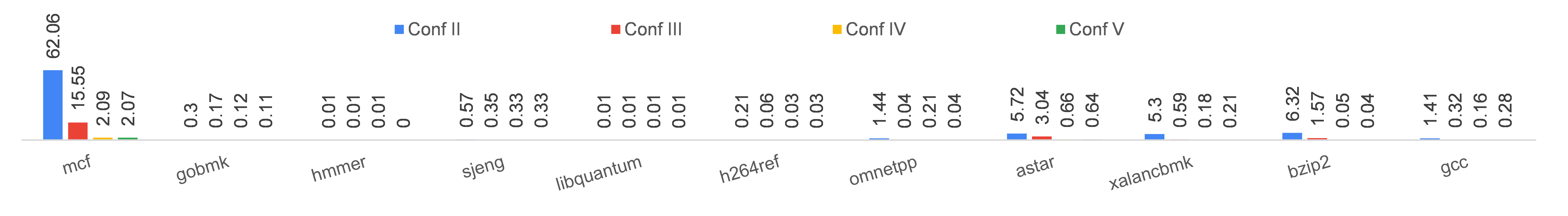} 
\vspace{-9pt}
\caption{Aggregated MPKI of the L2 TLB for the various TLB configurations.}
\vspace{-5pt}
\label{figure_l2_mpki}
\end{figure*}

Figure \ref{figure_l2_mpki} shows the results of the MPKI in the L2 TLB for the various configurations. The Configuration I is not included, as it lacks an L2 TLB. We observe that the L2 TLB MPKI for most benchmarks is nearly zero, particularly for the larger Configurations IV and V, thanks to the larger reach of the L2 TLB. There is also a major improvement in mcf which stresses the most the L2 TLB. On average, the miss rate for the L2 TLB is nearly zero with the larger Configurations IV and V.  

Focusing on the impact of associativity, 
Table~\ref{tableMissReductionL2} shows the number of L2 TLB misses for mcf as we increase the L2 TLB associativity but keep the number of L2 TLB entries constant.
The L1 Instruction/Data TLB parameters are based on those of Configuration V.

We observe that there is an 82.8\%/83.3\% reduction in TLB misses when associativity changes from direct-mapped to 4-way/8-way. This behavior highlights the possible impact on the miss rate that a direct-mapped TLB can have due to conflicting entries, and the benefits of using a set-associative TLB. Note, however, that such behavior depends on the working set of the application and its access pattern, and that our results are for the Spec2006int benchmarks with the rather small test input set, as explained in Section~\ref{sec:methodology}.

\begin{table}[ht]
\centering
\begin{tabular}{ l c c c } 
\midrule
\textbf{L2 TLB Associativity} & Direct-mapped & 4-way & 8-way \\
\textbf{\#TLB Misses for mcf}  & 40.2M  & 6.9M & 6.7M \\
\midrule
\end{tabular}
\caption{Number of L2 TLB misses for mcf as L2 TLB associativity increases, with Conf. V and fixed 1024-entry L2 TLB.}
\label{tableMissReductionL2}
\vspace{-17pt}
\end{table}

Finally, Table \ref{table_ipc_increase} summarizes the absolute IPC value with Configuration I, and the IPC speedup for the configurations II-V with respect to Configuration I. As we can see the IPC performance increases by up to 15.4 \% depending on the demand of TLB resources and access patterns of every benchmark.

\begin{table}[]
\centering
\begin{tabular}{ l c c c c c } 
\midrule
\textbf{Benchmark} & \textbf{I} & \textbf{II} & \textbf{III} & \textbf{IV} & \textbf{V}    \\ 
\midrule
mcf        & 0.13 & -      & 7.7 \% & 15.4 \% & 15.4 \%  \\ 
gobmk      & 0.44 & -      & -      & 2.3 \%  & 2.3 \%   \\
hmmer      & 0.58 & -      & -      & -       & -        \\
sjeng      & 0.55 & 1.8 \% & 1.8 \% & 1.8 \%  & 3.6 \%   \\
libquantum & 0.44 & -      & -      & -       & -        \\
h264ref    & 0.77 & 1.4 \% & 1.4 \% & 2.6 \%  & 2.6 \%   \\
omnetpp    & 0.35 & 2.9 \% & 5.7 \% & 5.7 \%  & 5.7 \%   \\
astar      & 0.36 & -      & -      & 2.8 \%  & 2.8 \%   \\
xalancbmk  & 0.36 & 2.8 \% & 8.3 \% & 8.3 \%  & 8.3 \%   \\
bzip2      & 0.51 & 2.0 \% & 4.0 \% & 5.9 \%  & 5.9 \%   \\
gcc        & 0.44 & 2.2 \% & 2.2 \% & 4.5 \%  & 4.5 \%   \\
\midrule
\end{tabular}
\caption{Absolute IPC values for Conf. I and percentage of IPC increase for Conf. II to V with respect to Conf. I.}
\label{table_ipc_increase}
\vspace{-25pt}
\end{table}
\section{Related Work}
\label{sec:related}

Prior work has focused on developing new MMU features for the Rocket Chip Generator in order to improve performance (e.g., Direct Segments for RISC-V \cite{kunati2018implementation})
while future work could investigate alternative techniques (e.g., Coalesced \cite{Pham2012} and Clustered TLBs \cite{pham}, Redundant Memory Mappings \cite{karakostas}, and Hybrid TLB Coalescing \cite{park}) to enhance the MMU performance.
Another line of prior work has focused on bridging the
FPGA-to-ASIC performance in order to gain more insights about the actual performance of a processor (e.g., \cite{biancolin, magyar, Kim2017EvaluationOR}) to be fabricated and to also lower resource usage. 
Furthermore, 
Content-Addressable-Memories (CAMs) are known to be resource-hungry structures \cite{wong}. Magyar et al. proposed Golden Gate \cite{magyar} to create Decoupled FPGA-accelerated Simulators by replacing FPGA-hostile CAMs with multi-cycle models, thus reducing resource utilization. As fully associative TLBs are typically implemented as CAMs, future work on resource optimization for large fully associative TLB organizations could leverage such FPGA-simulated research frameworks.

\section{Conclusions}
\label{sec:conclusions}

In this paper we explored the Memory Management Unit of the Rocket Chip Generator and lifted its implementation limitations in the TLB hierarchy. We implemented a fully configurable L1 and L2 TLB, that can output any design from direct-mapped to fully-associative. Our design enables design space exploration and allows the Rocket Chip Generator to instantiate cores with TLBs that match the needs of TLB intensive applications. We make our design publicly available to enable further research on the active topic of virtual memory support for the RISC-V architecture.

\section*{Acknowledgments}
We would like to thank Dr. Tuo Li from the UNSW School of Computer Science and Engineering  for porting the Rocket Chip Generator to the Xilinx ZCU102 board, his contribution and useful tips helped us considerably.

\balance
\bibliographystyle{ACM-Reference-Format}
\bibliography{sample-base}


\begin{thebibliography}{26}


\ifx \showCODEN    \undefined \def \showCODEN     #1{\unskip}     \fi
\ifx \showDOI      \undefined \def \showDOI       #1{#1}\fi
\ifx \showISBNx    \undefined \def \showISBNx     #1{\unskip}     \fi
\ifx \showISBNxiii \undefined \def \showISBNxiii  #1{\unskip}     \fi
\ifx \showISSN     \undefined \def \showISSN      #1{\unskip}     \fi
\ifx \showLCCN     \undefined \def \showLCCN      #1{\unskip}     \fi
\ifx \shownote     \undefined \def \shownote      #1{#1}          \fi
\ifx \showarticletitle \undefined \def \showarticletitle #1{#1}   \fi
\ifx \showURL      \undefined \def \showURL       {\relax}        \fi
\providecommand\bibfield[2]{#2}
\providecommand\bibinfo[2]{#2}
\providecommand\natexlab[1]{#1}
\providecommand\showeprint[2][]{arXiv:#2}

\bibitem[\protect\citeauthoryear{{A. Waterman and K. Asanovic}}{{A. Waterman
  and K. Asanovic}}{2017}]%
        {waterman02}
\bibfield{author}{\bibinfo{person}{{A. Waterman and K. Asanovic}}.}
  \bibinfo{year}{2017}\natexlab{}.
\newblock \bibinfo{title}{{The RISC-V Instruction Set Manual, Volume II:
  Privileged Architecture, Document Version 1.10}}.
\newblock
\newblock


\bibitem[\protect\citeauthoryear{{Andrew Waterman and K. Asanovic}}{{Andrew
  Waterman and K. Asanovic}}{2017}]%
        {waterman01}
\bibfield{author}{\bibinfo{person}{{Andrew Waterman and K. Asanovic}}.}
  \bibinfo{year}{2017}\natexlab{}.
\newblock \bibinfo{title}{{The RISC-V Instruction Set Manual, Volume I:
  User-Level ISA, Document Version 2.2}}.
\newblock
\newblock


\bibitem[\protect\citeauthoryear{{ARM}}{{ARM}}{[n.d.]}]%
        {arm-cortex-a57}
\bibfield{author}{\bibinfo{person}{{ARM}}.} \bibinfo{year}{[n.d.]}\natexlab{}.
\newblock \bibinfo{title}{{ARM Cortex-A57 Technical Reference Manual}}.
\newblock
\newblock
\urldef\tempurl%
\url{http://infocenter.arm.com/help/topic/com.arm.doc.ddi0488c/
  DDI0488C_cortex_a57_mpcore_r1p0_trm.pdf}
\showURL{%
\tempurl}


\bibitem[\protect\citeauthoryear{Asanović, Avizienis, Bachrach, Beamer,
  Biancolin, Celio, Cook, Dabbelt, Hauser, Izraelevitz, Karandikar, Keller,
  Kim, Koenig, Lee, Love, Maas, Magyar, Mao, Moreto, Ou, Patterson, Richards,
  Schmidt, Twigg, Vo, and Waterman}{Asanović et~al\mbox{.}}{2016}]%
        {asanovic01}
\bibfield{author}{\bibinfo{person}{Krste Asanović}, \bibinfo{person}{Rimas
  Avizienis}, \bibinfo{person}{Jonathan Bachrach}, \bibinfo{person}{Scott
  Beamer}, \bibinfo{person}{David Biancolin}, \bibinfo{person}{Christopher
  Celio}, \bibinfo{person}{Henry Cook}, \bibinfo{person}{Daniel Dabbelt},
  \bibinfo{person}{John Hauser}, \bibinfo{person}{Adam Izraelevitz},
  \bibinfo{person}{Sagar Karandikar}, \bibinfo{person}{Ben Keller},
  \bibinfo{person}{Donggyu Kim}, \bibinfo{person}{John Koenig},
  \bibinfo{person}{Yunsup Lee}, \bibinfo{person}{Eric Love},
  \bibinfo{person}{Martin Maas}, \bibinfo{person}{Albert Magyar},
  \bibinfo{person}{Howard Mao}, \bibinfo{person}{Miquel Moreto},
  \bibinfo{person}{Albert Ou}, \bibinfo{person}{David~A. Patterson},
  \bibinfo{person}{Brian Richards}, \bibinfo{person}{Colin Schmidt},
  \bibinfo{person}{Stephen Twigg}, \bibinfo{person}{Huy Vo}, {and}
  \bibinfo{person}{Andrew Waterman}.} \bibinfo{year}{2016}\natexlab{}.
\newblock \bibinfo{booktitle}{\emph{{The Rocket Chip Generator}}}.
\newblock \bibinfo{type}{{T}echnical {R}eport} UCB/EECS-2016-17.
  \bibinfo{institution}{EECS Department, University of California, Berkeley}.
\newblock
\urldef\tempurl%
\url{http://www2.eecs.berkeley.edu/Pubs/TechRpts/2016/EECS-2016-17.html}
\showURL{%
\tempurl}


\bibitem[\protect\citeauthoryear{Bachrach, Vo, Richards, Lee, Waterman,
  Avi\v{z}ienis, Wawrzynek, and Asanovi\'{c}}{Bachrach et~al\mbox{.}}{2012}]%
        {bachrach}
\bibfield{author}{\bibinfo{person}{Jonathan Bachrach}, \bibinfo{person}{Huy
  Vo}, \bibinfo{person}{Brian Richards}, \bibinfo{person}{Yunsup Lee},
  \bibinfo{person}{Andrew Waterman}, \bibinfo{person}{Rimas Avi\v{z}ienis},
  \bibinfo{person}{John Wawrzynek}, {and} \bibinfo{person}{Krste
  Asanovi\'{c}}.} \bibinfo{year}{2012}\natexlab{}.
\newblock \showarticletitle{Chisel: Constructing Hardware in a Scala Embedded
  Language}. In \bibinfo{booktitle}{\emph{Proceedings of the 49th Annual Design
  Automation Conference}} \emph{(\bibinfo{series}{DAC ’12})}.
  \bibinfo{publisher}{Association for Computing Machinery},
  \bibinfo{address}{New York, NY, USA}, \bibinfo{pages}{1216–1225}.
\newblock
\showISBNx{9781450311991}
\urldef\tempurl%
\url{https://doi.org/10.1145/2228360.2228584}
\showDOI{\tempurl}


\bibitem[\protect\citeauthoryear{Biancolin, Karandikar, Kim, Koenig, Waterman,
  Bachrach, and Asanovic}{Biancolin et~al\mbox{.}}{2019}]%
        {biancolin}
\bibfield{author}{\bibinfo{person}{David Biancolin}, \bibinfo{person}{Sagar
  Karandikar}, \bibinfo{person}{Donggyu Kim}, \bibinfo{person}{Jack Koenig},
  \bibinfo{person}{Andrew Waterman}, \bibinfo{person}{Jonathan Bachrach}, {and}
  \bibinfo{person}{Krste Asanovic}.} \bibinfo{year}{2019}\natexlab{}.
\newblock \showarticletitle{FASED: FPGA-Accelerated Simulation and Evaluation
  of DRAM}. In \bibinfo{booktitle}{\emph{Proceedings of the 2019 ACM/SIGDA
  International Symposium on Field-Programmable Gate Arrays}}
  \emph{(\bibinfo{series}{FPGA ’19})}. \bibinfo{publisher}{Association for
  Computing Machinery}, \bibinfo{address}{New York, NY, USA},
  \bibinfo{pages}{330–339}.
\newblock
\showISBNx{9781450361378}
\urldef\tempurl%
\url{https://doi.org/10.1145/3289602.3293894}
\showDOI{\tempurl}


\bibitem[\protect\citeauthoryear{{Buildroot}}{{Buildroot}}{2017}]%
        {buildroot}
\bibfield{author}{\bibinfo{person}{{Buildroot}}.}
  \bibinfo{year}{{2017}}\natexlab{}.
\newblock \bibinfo{title}{{Buildroot manual}}.
\newblock
\newblock
\urldef\tempurl%
\url{https://buildroot.org/downloads/manual/manual.html}
\showURL{%
\tempurl}


\bibitem[\protect\citeauthoryear{Celio, Patterson, and Asanović}{Celio
  et~al\mbox{.}}{2015}]%
        {Celio:EECS-2015-167}
\bibfield{author}{\bibinfo{person}{Christopher Celio},
  \bibinfo{person}{David~A. Patterson}, {and} \bibinfo{person}{Krste
  Asanović}.} \bibinfo{year}{2015}\natexlab{}.
\newblock \bibinfo{booktitle}{\emph{{The Berkeley Out-of-Order Machine (BOOM):
  An Industry-Competitive, Synthesizable, Parameterized RISC-V Processor}}}.
\newblock \bibinfo{type}{{T}echnical {R}eport} UCB/EECS-2015-167.
  \bibinfo{institution}{EECS Department, University of California, Berkeley}.
\newblock
\urldef\tempurl%
\url{http://www2.eecs.berkeley.edu/Pubs/TechRpts/2015/EECS-2015-167.html}
\showURL{%
\tempurl}


\bibitem[\protect\citeauthoryear{{Christopher Celio}}{{Christopher
  Celio}}{[n.d.]}]%
        {celio02}
\bibfield{author}{\bibinfo{person}{{Christopher Celio}}.}
  \bibinfo{year}{[n.d.]}\natexlab{}.
\newblock \bibinfo{title}{{Speckle: A wrapper for the SPEC CPU2006 benchmark
  suite}}.
\newblock
\newblock
\urldef\tempurl%
\url{https://github.com/ccelio/Speckle}
\showURL{%
\tempurl}


\bibitem[\protect\citeauthoryear{Henning}{Henning}{2006}]%
        {spec01}
\bibfield{author}{\bibinfo{person}{John~L. Henning}.}
  \bibinfo{year}{2006}\natexlab{}.
\newblock \showarticletitle{SPEC CPU2006 Benchmark Descriptions}.
\newblock \bibinfo{journal}{\emph{SIGARCH Comput. Archit. News}}
  \bibinfo{volume}{34}, \bibinfo{number}{4} (\bibinfo{date}{Sept.}
  \bibinfo{year}{2006}), \bibinfo{pages}{1–17}.
\newblock
\showISSN{0163-5964}
\urldef\tempurl%
\url{https://doi.org/10.1145/1186736.1186737}
\showDOI{\tempurl}


\bibitem[\protect\citeauthoryear{{Intel}}{{Intel}}{[n.d.]}]%
        {haswell}
\bibfield{author}{\bibinfo{person}{{Intel}}.}
  \bibinfo{year}{[n.d.]}\natexlab{}.
\newblock \bibinfo{title}{{Intel® 64 and IA-32 Architectures Optimization
  Reference Manual}}.
\newblock
\newblock
\urldef\tempurl%
\url{https://www.intel.com/content/dam/www/public/us/en/documents/
  manuals/64-ia-32-architectures-optimization-manual.pdf}
\showURL{%
\tempurl}


\bibitem[\protect\citeauthoryear{{Karakostas}, {Gandhi}, {Ayar}, {Cristal},
  {Hill}, {McKinley}, {Nemirovsky}, {Swift}, and {Ünsal}}{{Karakostas}
  et~al\mbox{.}}{2015}]%
        {karakostas}
\bibfield{author}{\bibinfo{person}{V. {Karakostas}}, \bibinfo{person}{J.
  {Gandhi}}, \bibinfo{person}{F. {Ayar}}, \bibinfo{person}{A. {Cristal}},
  \bibinfo{person}{M.~D. {Hill}}, \bibinfo{person}{K.~S. {McKinley}},
  \bibinfo{person}{M. {Nemirovsky}}, \bibinfo{person}{M.~M. {Swift}}, {and}
  \bibinfo{person}{O. {Ünsal}}.} \bibinfo{year}{2015}\natexlab{}.
\newblock \showarticletitle{Redundant Memory Mappings for fast access to large
  memories}. In \bibinfo{booktitle}{\emph{2015 ACM/IEEE 42nd Annual
  International Symposium on Computer Architecture (ISCA)}}.
  \bibinfo{pages}{66--78}.
\newblock


\bibitem[\protect\citeauthoryear{Kim, Celio, Biancolin, Bachrach, and
  Asanovic}{Kim et~al\mbox{.}}{2017}]%
        {Kim2017EvaluationOR}
\bibfield{author}{\bibinfo{person}{Donggyu Kim}, \bibinfo{person}{Christopher
  Celio}, \bibinfo{person}{David Biancolin}, \bibinfo{person}{Jonathan
  Bachrach}, {and} \bibinfo{person}{Krste Asanovic}.}
  \bibinfo{year}{2017}\natexlab{}.
\newblock \showarticletitle{Evaluation of RISC-V RTL with FPGA-Accelerated
  Simulation}.
\newblock


\bibitem[\protect\citeauthoryear{Kunati and Swift}{Kunati and Swift}{2018}]%
        {kunati2018implementation}
\bibfield{author}{\bibinfo{person}{Nikhita Kunati} {and}
  \bibinfo{person}{Michael~M Swift}.} \bibinfo{year}{2018}\natexlab{}.
\newblock \showarticletitle{Implementation of Direct Segments on a RISC-V
  Processor}.
\newblock  (\bibinfo{year}{2018}).
\newblock


\bibitem[\protect\citeauthoryear{{Magyar}, {Biancolin}, {Koenig}, {Seshia},
  {Bachrach}, and {Asanović}}{{Magyar} et~al\mbox{.}}{2019}]%
        {magyar}
\bibfield{author}{\bibinfo{person}{A. {Magyar}}, \bibinfo{person}{D.
  {Biancolin}}, \bibinfo{person}{J. {Koenig}}, \bibinfo{person}{S. {Seshia}},
  \bibinfo{person}{J. {Bachrach}}, {and} \bibinfo{person}{K. {Asanović}}.}
  \bibinfo{year}{2019}\natexlab{}.
\newblock \showarticletitle{Golden Gate: Bridging The Resource-Efficiency Gap
  Between ASICs and FPGA Prototypes}. In \bibinfo{booktitle}{\emph{2019
  IEEE/ACM International Conference on Computer-Aided Design (ICCAD)}}.
  \bibinfo{pages}{1--8}.
\newblock


\bibitem[\protect\citeauthoryear{Park, Heo, Jeong, and Huh}{Park
  et~al\mbox{.}}{2017}]%
        {park}
\bibfield{author}{\bibinfo{person}{Chang~Hyun Park}, \bibinfo{person}{Taekyung
  Heo}, \bibinfo{person}{Jungi Jeong}, {and} \bibinfo{person}{Jaehyuk Huh}.}
  \bibinfo{year}{2017}\natexlab{}.
\newblock \showarticletitle{Hybrid TLB Coalescing: Improving TLB Translation
  Coverage under Diverse Fragmented Memory Allocations}. In
  \bibinfo{booktitle}{\emph{Proceedings of the 44th Annual International
  Symposium on Computer Architecture}} \emph{(\bibinfo{series}{ISCA ’17})}.
  \bibinfo{publisher}{Association for Computing Machinery},
  \bibinfo{address}{New York, NY, USA}, \bibinfo{pages}{444–456}.
\newblock
\showISBNx{9781450348928}
\urldef\tempurl%
\url{https://doi.org/10.1145/3079856.3080217}
\showDOI{\tempurl}


\bibitem[\protect\citeauthoryear{{Pham}, {Bhattacharjee}, {Eckert}, and
  {Loh}}{{Pham} et~al\mbox{.}}{2014}]%
        {pham}
\bibfield{author}{\bibinfo{person}{B. {Pham}}, \bibinfo{person}{A.
  {Bhattacharjee}}, \bibinfo{person}{Y. {Eckert}}, {and} \bibinfo{person}{G.~H.
  {Loh}}.} \bibinfo{year}{2014}\natexlab{}.
\newblock \showarticletitle{Increasing TLB reach by exploiting clustering in
  page translations}. In \bibinfo{booktitle}{\emph{2014 IEEE 20th International
  Symposium on High Performance Computer Architecture (HPCA)}}.
  \bibinfo{pages}{558--567}.
\newblock


\bibitem[\protect\citeauthoryear{Pham, Vaidyanathan, Jaleel, and
  Bhattacharjee}{Pham et~al\mbox{.}}{2012}]%
        {Pham2012}
\bibfield{author}{\bibinfo{person}{Binh Pham}, \bibinfo{person}{Viswanathan
  Vaidyanathan}, \bibinfo{person}{Aamer Jaleel}, {and}
  \bibinfo{person}{Abhishek Bhattacharjee}.} \bibinfo{year}{2012}\natexlab{}.
\newblock \showarticletitle{CoLT: Coalesced Large-Reach TLBs}. In
  \bibinfo{booktitle}{\emph{Proceedings of the 2012 45th Annual IEEE/ACM
  International Symposium on Microarchitecture}}
  \emph{(\bibinfo{series}{MICRO-45})}. \bibinfo{publisher}{IEEE Computer
  Society}, \bibinfo{address}{USA}, \bibinfo{pages}{258–269}.
\newblock
\showISBNx{9780769549248}
\urldef\tempurl%
\url{https://doi.org/10.1109/MICRO.2012.32}
\showDOI{\tempurl}


\bibitem[\protect\citeauthoryear{{RISC-V Foundation}}{{RISC-V
  Foundation}}{[n.d.]a}]%
        {riscv-pk}
\bibfield{author}{\bibinfo{person}{{RISC-V Foundation}}.}
  \bibinfo{year}{[n.d.]}\natexlab{a}.
\newblock \bibinfo{title}{{riscv-pk}}.
\newblock
\newblock
\urldef\tempurl%
\url{https://github.com/riscv/riscv-pk/tree/master/bbl}
\showURL{%
\tempurl}


\bibitem[\protect\citeauthoryear{{RISC-V Foundation}}{{RISC-V
  Foundation}}{[n.d.]b}]%
        {riscv-tests}
\bibfield{author}{\bibinfo{person}{{RISC-V Foundation}}.}
  \bibinfo{year}{[n.d.]}\natexlab{b}.
\newblock \bibinfo{title}{{riscv-tests}}.
\newblock
\newblock
\urldef\tempurl%
\url{https://github.com/riscv/riscv-tests}
\showURL{%
\tempurl}


\bibitem[\protect\citeauthoryear{{Sifive}}{{Sifive}}{[n.d.]}]%
        {sifive01}
\bibfield{author}{\bibinfo{person}{{Sifive}}.}
  \bibinfo{year}{[n.d.]}\natexlab{}.
\newblock \bibinfo{title}{{Freedom-U-SDK}}.
\newblock
\newblock
\urldef\tempurl%
\url{https://github.com/sifive/freedom-u-sdk}
\showURL{%
\tempurl}


\bibitem[\protect\citeauthoryear{Talluri and Hill}{Talluri and Hill}{1994}]%
        {talluri02}
\bibfield{author}{\bibinfo{person}{Madhusudhan Talluri} {and}
  \bibinfo{person}{Mark~D. Hill}.} \bibinfo{year}{1994}\natexlab{}.
\newblock \showarticletitle{Surpassing the TLB Performance of Superpages with
  Less Operating System Support}. In \bibinfo{booktitle}{\emph{Proceedings of
  the Sixth International Conference on Architectural Support for Programming
  Languages and Operating Systems}} \emph{(\bibinfo{series}{ASPLOS VI})}.
  \bibinfo{publisher}{Association for Computing Machinery},
  \bibinfo{address}{New York, NY, USA}, \bibinfo{pages}{171–182}.
\newblock
\showISBNx{0897916603}
\urldef\tempurl%
\url{https://doi.org/10.1145/195473.195531}
\showDOI{\tempurl}


\bibitem[\protect\citeauthoryear{Talluri, Kong, Hill, and Patterson}{Talluri
  et~al\mbox{.}}{1992}]%
        {talluri01}
\bibfield{author}{\bibinfo{person}{Madhusudhan Talluri}, \bibinfo{person}{Shing
  Kong}, \bibinfo{person}{Mark~D. Hill}, {and} \bibinfo{person}{David~A.
  Patterson}.} \bibinfo{year}{1992}\natexlab{}.
\newblock \showarticletitle{Tradeoffs in Supporting Two Page Sizes}. In
  \bibinfo{booktitle}{\emph{Proceedings of the 19th Annual International
  Symposium on Computer Architecture}} \emph{(\bibinfo{series}{ISCA ’92})}.
  \bibinfo{publisher}{Association for Computing Machinery},
  \bibinfo{address}{New York, NY, USA}, \bibinfo{pages}{415–424}.
\newblock
\showISBNx{0897915097}
\urldef\tempurl%
\url{https://doi.org/10.1145/139669.140406}
\showDOI{\tempurl}


\bibitem[\protect\citeauthoryear{{Wilson Snyder}}{{Wilson Snyder}}{[n.d.]}]%
        {verilator}
\bibfield{author}{\bibinfo{person}{{Wilson Snyder}}.}
  \bibinfo{year}{[n.d.]}\natexlab{}.
\newblock \bibinfo{title}{{Verilator Manual}}.
\newblock
\newblock
\urldef\tempurl%
\url{https://www.veripool.org/wiki/verilator/Manual-verilator}
\showURL{%
\tempurl}


\bibitem[\protect\citeauthoryear{{Wong}, {Betz}, and {Rose}}{{Wong}
  et~al\mbox{.}}{2014a}]%
        {wong01}
\bibfield{author}{\bibinfo{person}{H. {Wong}}, \bibinfo{person}{V. {Betz}},
  {and} \bibinfo{person}{J. {Rose}}.} \bibinfo{year}{2014}\natexlab{a}.
\newblock \showarticletitle{Quantifying the Gap Between FPGA and Custom CMOS to
  Aid Microarchitectural Design}.
\newblock \bibinfo{journal}{\emph{IEEE Transactions on Very Large Scale
  Integration (VLSI) Systems}} \bibinfo{volume}{22}, \bibinfo{number}{10}
  (\bibinfo{year}{2014}), \bibinfo{pages}{2067--2080}.
\newblock


\bibitem[\protect\citeauthoryear{{Wong}, {Betz}, and {Rose}}{{Wong}
  et~al\mbox{.}}{2014b}]%
        {wong}
\bibfield{author}{\bibinfo{person}{H. {Wong}}, \bibinfo{person}{V. {Betz}},
  {and} \bibinfo{person}{J. {Rose}}.} \bibinfo{year}{2014}\natexlab{b}.
\newblock \showarticletitle{Quantifying the Gap Between FPGA and Custom CMOS to
  Aid Microarchitectural Design}.
\newblock \bibinfo{journal}{\emph{IEEE Transactions on Very Large Scale
  Integration (VLSI) Systems}} \bibinfo{volume}{22}, \bibinfo{number}{10}
  (\bibinfo{year}{2014}), \bibinfo{pages}{2067--2080}.
\newblock


\end{thebibliography}
\end{document}